# Large-scale structure and motions from simulated galaxy clusters.

R. A. C. Croft and G. Efstathiou *

**Abstract**

We use high resolution dissipationless N-body simulations to examine the spatial distribution of galaxy clusters on large scales. The Standard CDM model and two of its main competitors, Low density CDM and Mixed Dark Matter are compared. The two-point correlation function of simulated clusters is compared with an extended survey of APM clusters , and it is found that Standard CDM exhibits a lack of power on *all* scales, whereas the two alternative scenarios are able to match the spatial correlations well. Of the remaining two models, the velocities in the MDM universe have a higher amplitude and their distribution is much broader . We compare these peculiar velocities with observations and find that both models have difficulty in reproducing the observed numbers of very high peculiar velocity clusters. The reliable detection of several more clusters with velocities in excess of 1000km s$^{-1}$ would render the LCDM scenario in particular very unlikely.

## 1. Introduction

Clusters of galaxies efficiently trace out the large scale structure of the universe. Measurements of their spatial correlation function in and 2 and 3 dimensions[11,3,13] provided some of the first evidence that there is more power in the clustering of matter than can be accounted for by the standard CDM model. More recently, surveys of rich clusters picked from computer generated galaxy catalogues[7,17] , and a survey of ROSAT X-ray clusters[18] indicate that their 3D two-point correlation function has the following form:

$$\xi_{cc}(r) \approx (r/r_0)^{-1.8} \qquad r_0 \approx 13 - 16 \ h^{-1} \mathrm{Mpc} \qquad (1)$$

(the Hubble constant is $H_0 = 100h\mathrm{km\ s^{-1}\ Mpc^{-1}}$). We have run a series of N-body simulations to determine whether equation (1) can be reproduced by the standard CDM model, or by variants such as Low density CDM or Mixed Dark Matter. The simulations, described in Section 2.1 , are large enough to resolve individual clusters and sample a large enough volume of space to accurately determine the cluster two-point correlation function.

Measurements of a coherent large-scale velocity field[10] have also been cited as evidence against the standard model. Galaxy clusters play an important role in these measurements too. The observational errors on galaxy velocities can be large. However the distance errors to rich clusters can be reduced by $1/\sqrt{N}$ where $N$ is the number of cluster galaxies with independent distance measurements. Here we compare cluster peculiar velocities with those measured from N-body simulations.

---

*Department of Physics, University of Oxford, Keble Road, Oxford, OX1 3RH, U. K. This work was supported by the UK SERC.

## 2. Simulating cluster formation

### 2.1. N-body simulations

The CDM-like N-body models are described in detail in ref. [5]. A P$^3$M code [9] is used to follow the evolution of the dissipationless component of the mass. We simulate a box of side 300 $h^{-1}$Mpc, with $10^6$ particles per simulation and a spatial resolution of $80h^{-1}$kpc. For each cosmological model, we make 10 runs, using different random phases to generate the initial conditions. We simulate spatially flat universes, with Gaussian initial conditions and the initial power spectrum of mass fluctuations derived from linear theory calculations of the evolution of adiabatic fluctuations[12,19]. The three sets of models are as follows:

(a) Standard CDM, with $\Omega = 1$ and $h = 0.5$.

(b) Low density CDM ( hereafter LCDM), with $\Omega = 0.2, h = 1.0$ and a non-zero cosmological constant, $\Lambda = 0.8(3H_0^2)$ introduced to make the model spatially flat.

(c) Mixed dark matter (hereafter MDM) with 7ev neutrinos: $\Omega_\nu = 0.3$, $\quad \Omega_{CDM} = 0.6$.

For model (c), the power spectrum of total mass is used[12], but the effect of neutrino thermal velocities is not included, as here we study only the large-scale distribution of clusters and not their detailed internal structure. To find the 'present day', the abundance of rich clusters was fixed to be its observed value[19] . This criterion, together with a best fit to the COBE data requires that the linear amplitude of mass fluctuations in 8 $h^{-1}$Mpc spheres ($\sigma_8$) should be $\simeq 0.57$ for the $\Omega = 1$ models and $\simeq 1.0$ for LCDM. It will be shown in Section (3) that the spatial correlations are insensitive to changes in the value of $\sigma_8$.

### 2.2. Cluster selection

We find clusters by computing the mass contained within non-overlapping spheres of radius r$_c$ centred on mass concentrations found by a percolation algorithm. In [5] we show that the catalogue of clusters picked out is insensitive to variations in r$_c$ (here we use 0.5 $h^{-1}$Mpc). By a applying a lower mass bound we select clusters with the same space density as the observational sample we are comparing with, in our case the extended APM cluster redshift survey[7] (mean intercluster separation=30.5 $h^{-1}$Mpc). As long as there is a roughly monotonic relationship between the mass of a cluster and its luminosity, then we can bypass any other assumptions concerning assignment of galaxies to the clusters.

## 3. Cluster spatial correlations

The two point correlation function of these clusters is shown in Figure (1) for our three different cosmologies, as well as the observational points from the extended APM cluster survey. This survey consists of 364 clusters picked from the APM galaxy survey. The error bars are small, and it can be seen that $\xi_{cc}$ for Standard CDM has too low an amplitude at all separations, unlike the other two models which fit the correlations much better over the full range $2 \leq r \leq 40$ $h^{-1}$Mpc. Varying the degree of evolution of the mass density field ($\sigma_8$) does not significantly alter the results. The thick lines on each plot are $\xi_{cc}$ in redshift space from simulated APM-style catalogues, using the APM survey mask and selection function (see [5] for details). This does not really alter the results, save a slight tilt which brings models (b) and (c) into even better agreement with the data. Spatial modulation of the galaxy distribution by quasars[2] or 'cooperative' effects[4] has been put forward as a potential way of reconciling Standard CDM with observations that show too much large scale power. In the case of these cluster correlations, such effects would also have to work, with a high amplitude, on scales where the mass distribution is strongly non-linear, and it seems unlikely that this is feasible. In the next section, we therefore choose to concentrate on differentiating the other between MDM and LCDM (the work is described in more detail in [6])

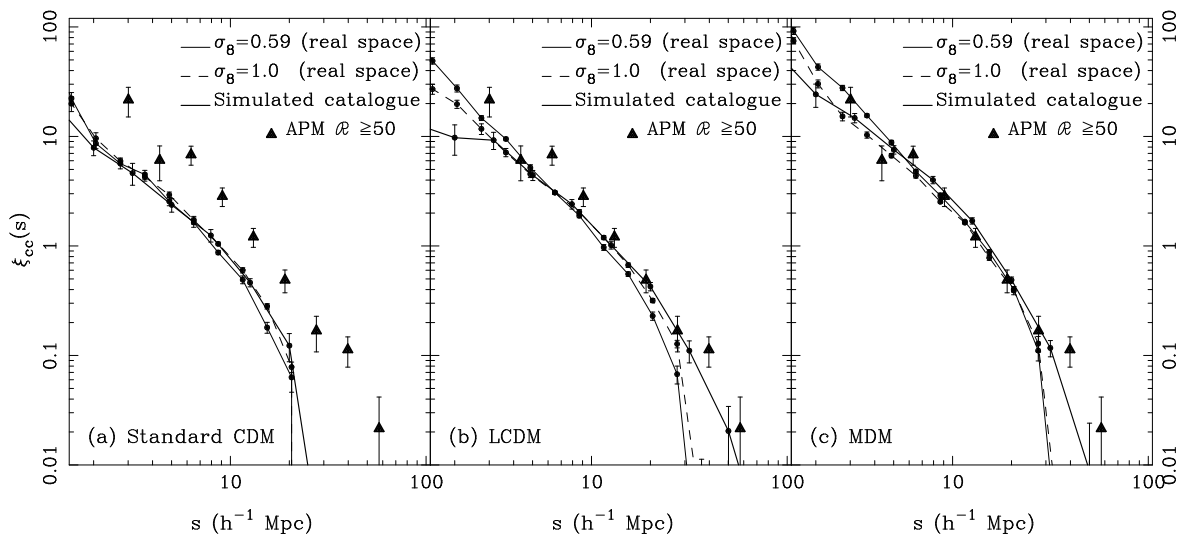

Figure 1. The two-point correlation function for simulated clusters with the same space density as APM $\mathcal{R} \geq 50$ clusters (mean intercluster distance = 30.5 $h^{-1}$Mpc) . $\xi_{cc}$ is plotted in real space for two values of the amplitude of mass fluctuations, $\sigma_8$ (thin lines). The error bars are derived from the scatter between 10 simulations for each ensemble. Also plotted (thick lines) is $\xi_{cc}$ in redshift space, using simulated APM-style catalogues (see text). The observed $\xi_{cc}$ calculated from the extended APM cluster redshift survey is shown, with Poissonian error bars.

## 4. Cluster peculiar velocities

The distribution of three dimensional peculiar velocities for clusters in the LCDM and MDM simulations is shown in Figure (2). That the distribution is much flatter, with a higher rms value (700km s$^{-1}$ as opposed to 400km s$^{-1}$) for MDM than LCDM. The histograms depend on the value of $\sigma_8$, so when comparing with observations, the curve corresponding to $\sigma_8$ given in Section (2) should be used.

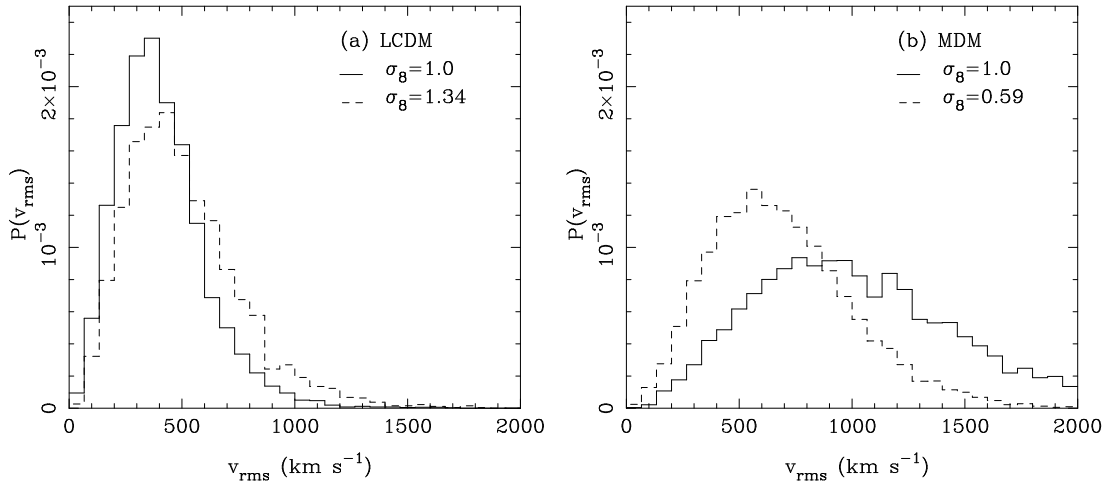

Figure 2.  The distribution of rms cluster peculiar velocities for two different values of $\sigma_8$ in simulations of (a) Low density CDM with a cosmological constant and (b) Mixed Dark Matter.

Observationally, the line-of-sight component of velocity is determined using both a redshift ($z$), and a distance indicator to give $v = H_0 r - cz$. Here we use an observational sample of 65 clusters in total - the distances were calculated using either the Tully-Fisher [1,14,15,16] or $D_n - \sigma$ [10] relations. We choose to exclude the 5 clusters with distance errors $\geq 850$km s$^{-1}$. To compare with observations, we broaden our simulated cluster velocity distributions with the observational errors by adding an error to each velocity at random from a Gaussian distribution with dispersion equal to that of the observational error. This process is repeated 100 times for each simulated cluster and the resulting distribution smoothed with a Gaussian filter of width 400km s$^{-1}$. In Figure (3) the probability of picking a cluster at random with a velocity greater than a certain value is shown. Finding clusters with high velocities is much more likely in the observational sample than in both models. However, the observational errors are certainly extremely large, and there may be important systematic effects - for example the $D_n - \sigma$ velocities are often larger than those determined using other methods. Another problem is the fact that the many of the observed clusters sample the flow to the Great Attractor - our 60 clusters are certainly not a 'fair sample'.

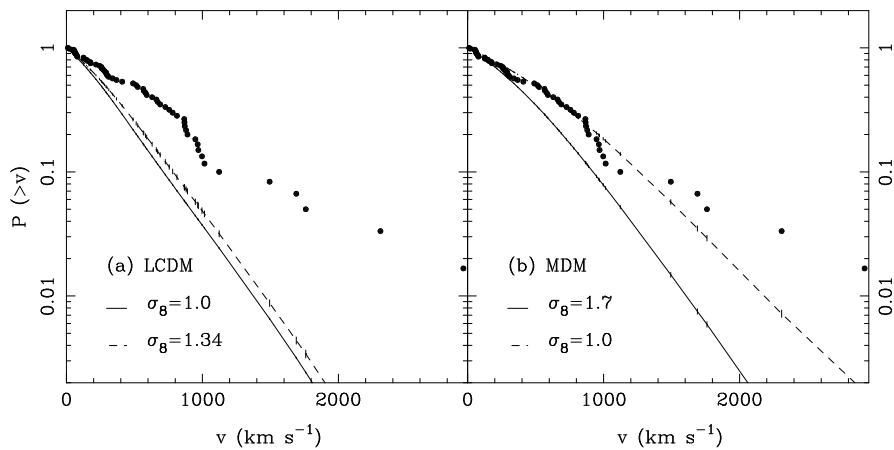

Figure 3. The cumulative probability distribution of cluster peculiar velocities. The points in each panel represent the observed sample (which is probably unrepresentative of the universe as a whole). The lines represent the values derived from the 10 numerical simulations for each model, which have been broadened with observational errors. (a) is Low density CDM and (b) Mixed Dark Matter.

## 5. Conclusions

The correlation function of rich clusters of galaxies is a controversial subject, but it appears that simulations of Standard CDM where clusters are selected in a similar way to observations are unable to match values obtained from real samples. Of the remaining two models that we test, observations of peculiar velocities appear to favour a model with a high proportion of clusters with large velocities, such as MDM. Neither model matches the results that well, though. Whilst the flows detected by surveys are almost certainly real, their magnitude is debatable. For the models to be correct, we must live in a region which more heavily samples the high velocity tail of the distribution. This is particularly true of LCDM, which could be ruled out if the observed values used in this study turn out to be true and representative of good sample of the universe.